%% file: cluster_pt.tex
\documentclass[sigconf]{acmart}

\setcitestyle{square}
\usepackage{booktabs} 
\usepackage[ruled]{algorithm2e} 
\usepackage{subcaption}
\usepackage{grffile}

\copyrightyear{2021}
\acmYear{2021}
\setcopyright{acmcopyright}\acmConference[SA '21 Technical Communications]{SIGGRAPH Asia 2021 Technical Communications}{December 14--17, 2021}{Tokyo, Japan}
\acmBooktitle{SIGGRAPH Asia 2021 Technical Communications (SA '21 Technical Communications), December 14--17, 2021, Tokyo, Japan}
\acmPrice{15.00}
\acmDOI{10.1145/3478512.3488605}
\acmISBN{978-1-4503-9073-6/21/12}

\hypersetup{draft}
\begin{document}

\title{Real Time Cluster Path Tracing}

\author{Feng Xie}
\affiliation{%
  \institution{Facebook Reality Labs}
}

\author{Petro Mishchuk}
\affiliation{%
  \institution{Apex Systems}
}

\author{Warren Hunt}
\affiliation{%
  \institution{Facebook Reality Labs}
}

\renewcommand{\shortauthors}{Xie et al.}
\newcommand{\omegaH} {\omega_{H}}
\newcommand{\brdf} {BRDF}
\newcommand{\bsdf} {BSDF}
\newcommand{\pdf} {PDF}
\newcommand{\fr} {\f_r(\omega_o, \omega_i)}
\newcommand{\J}{$J$}
\newcommand{\F}{$F$}

\begin{teaserfigure}
\vspace{-1.0em}
  \includegraphics[width=0.95\textwidth]{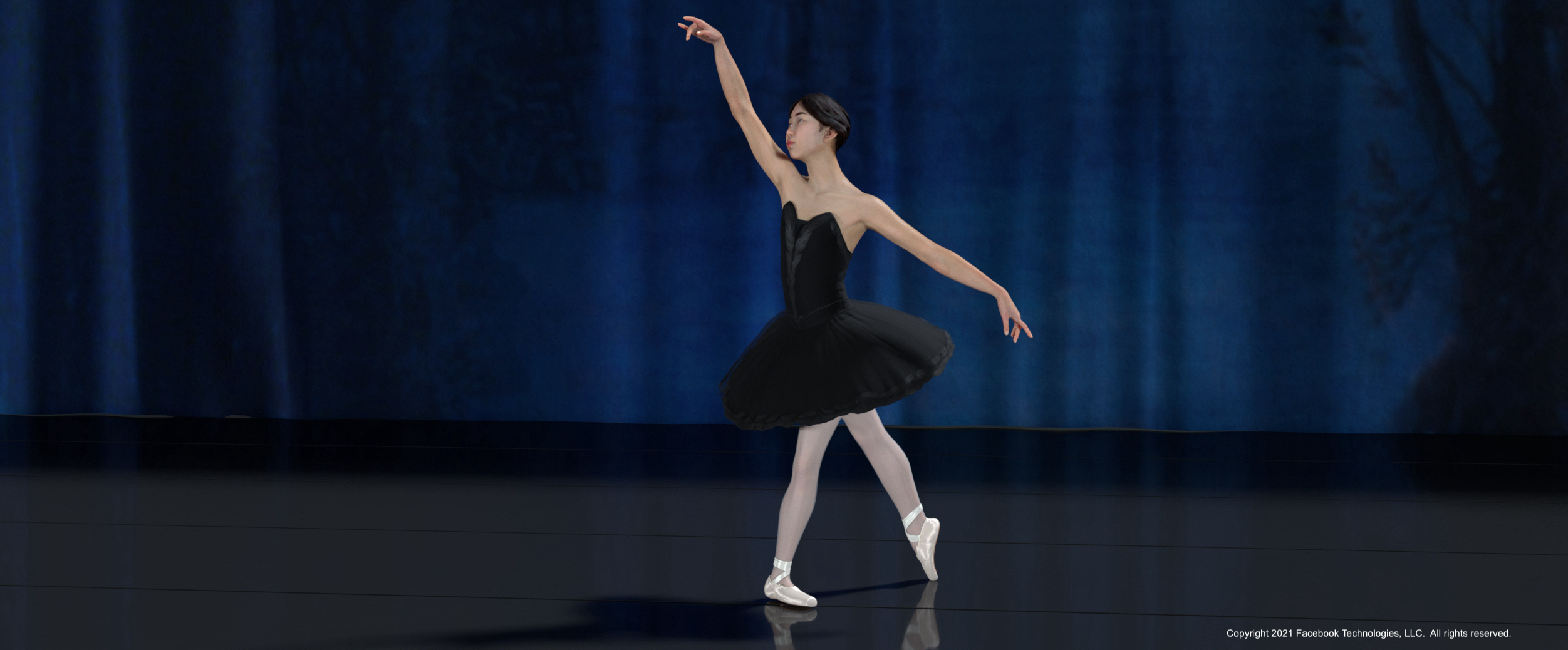}
\vspace{-0.5em}
  \caption{Trudy as Black Swan rendered with our cluster based photorealistic rendering system.}
  \label{fig:teaser}
\end{teaserfigure}

%
%
\ccsdesc[500]{Computing methodologies~Rendering}
\ccsdesc[500]{Computing methodologies~Raytracing}
\ccsdesc[500]{Computing methodologies~Distributed Algorithms}
\ccsdesc[500]{Computing methodologies~Parallel Algorithms}
\vspace{-1em}
\keywords{Real Time Path Tracing, Distributed Rendering, Scalability and Performance, Real Time Global Illumination}

\settopmatter{printfolios=true}
\maketitle

\input{introduction}

\input{related}

\input{cluster}

\input{realtime}

\input{results}
\vspace{-0.5em}

\paragraph{Acknowledgments}
Thanks to Brecht Van Lommel and Kevin Dietrich for
their work on the Cycles \emph{Alembic} Procedural.
Thanks to Fang Yu, Brian Budge, Mark Segal, and David Blythe
for their comments and suggestions on the paper revisions.
\vspace{-1em}
\bibliographystyle{ACM-Reference-Format}
\nocite{*}
\bibliography{cluster_pt}

\end{document}


\title{Real Time Cluster Path Tracing \\ Supplemental Material}

\author{Feng Xie}
\affiliation{%
  \institution{Facebook Reality Labs}
  \country{}
}

\author{Petro Mishchuk}
\affiliation{%
  \institution{Apex Systems}
}

\author{Warren Hunt}
\affiliation{%
  \institution{Facebook Reality Labs}
}

\renewcommand{\shortauthors}{Xie et al.}
\newcommand{\omegaH} {\omega_{H}}
\newcommand{\brdf} {BRDF}
\newcommand{\bsdf} {BSDF}
\newcommand{\pdf} {PDF}
\newcommand{\fr} {\f_r(\omega_o, \omega_i)}
\newcommand{\J}{$J$}
\newcommand{\F}{$F$}
\newcommand{\ignorethis}[1]{}

%
%
\maketitle
\input{supplemental_body}



%% file: introduction.tex
\vspace{-0.5em}
\section {Introduction}
Photorealistic rendering effects are common in films, but most real time 
graphics today still rely on scanline based 
multi-pass rendering to deliver rich visual experiences.

In this paper, we present the architecture and implementation of the first production quality real-time
cluster path tracing renderer.
We build our cluster path tracing system using the open source Blender 
and its GPU accelerated 
production quality renderer Cycles \cite{Cycles2021}.
Our system's rendering performance and quality scales linearly with the number of RTX 
cluster nodes used.
It is able to generate and deliver path traced images with global illumination effects
to remote light-weight client systems at $15-30$ frames per second for a wide variety of Blender scenes including virtual objects and 
animated digital human characters. 

%% file: related.tex
\vspace{-0.5em}
\section {Related Work}
The path tracing revolution in CG production started around 2007.
With the release of the film
\textit{Cloudy with a Chance of Meatballs}, Sony rocked production rendering by using the brute force
path tracer Arnold \cite{SonyArnold2018}.  Since then most major
CG film productions have moved onto path traced rendering.  
In 2018, 
TOG published a special issue on production path tracing renderers
including Maya \cite{Arnold2018} and Sony Arnold \cite{SonyArnold2018}, 
Weta's Manuka \cite{Manuka2018}, 
Disney's Hyperion \cite{Hyperion2018} and Pixar's Renderman \cite{RenderMan2018}.
These papers present 
the benefits of physically based rendering  w.r.t. artist 
workflow, image quality consistency and predictability.  
Path tracing isn't just able
to deliver physically based photorealism, 
but can also be used to create higly stylized films 
like Sony's \textit{Spiderman in the Spider-verse} and DreamWorks' furry \textit{Trolls}. 


For real time PC and console games, the launch of Nvidia's RTX GPUs and Intel's Xe Graphics, along with
the support for ray tracing APIs in Direct X and Vulcan translated to broader adoption of ray tracing
based reflections and shadows.
However, most games today still 
use scanline based rendering system as a foundation, 
with effects like global illumination, 
subsurface scattering, area lights, environment
lighting supported using a multi-pass or light baking approach.  
The Lumen global illumination system in Unreal Engine 5.0 \cite{UnrealLumen} 
supports 
software ray tracing for signed distance fields and hardware accelerated ray tracing for mesh geometry, 
with some limitations like no support for transparent materials in the former 
and constraints around complex deforming characters in the latter.  The system has no support
for cluster distribution of ray tracing for scalability.

We choose to build a real time cluster renderer using path tracing not only
because of its generality in terms of photorealistic rendering effects
but also because
it is uniquely suited for massively parallel computing, a quality that has been exploited
in the many works related to ray tracing acceleration.  
Wald et al. \shortcite{DistributedInteractive} present interactive
ray tracing for simple objects without any scattering effects on a cluster of PCs.
Benthin et al. \shortcite{ClusterGI} present a global illumination solution using
clusters with up for 48 CPUs  to 
achieve interactive frame rates at $2$ to $5$ fps for static scenes.  
Jaros et al.\shortcite{BlenderMPI2017} present using MPI to accelerate Blender Cycles on a Xeon Phi-based cluster,
and Gerveshi and Looper \shortcite{DistributedMoonray2019} present distributed interactive rendering using 
the Moonray \cite{Moonray2017}
renderer;
neither system supports real time animation.

While there has been much work done to accelerate and advance ray tracing and path tracing
across both offline and real time rendering, we are the first to
present a distributed path tracing rendering system that leverages cluster computing
and low latency streaming to deliver real time 
photorealistic rendering of complex scenes and dynamic characters
to consumer platforms without special GPU support.

%% file: cluster.tex
\begin{figure}[h]
\vspace{-0.5em}
\includegraphics[width=0.25\textwidth]{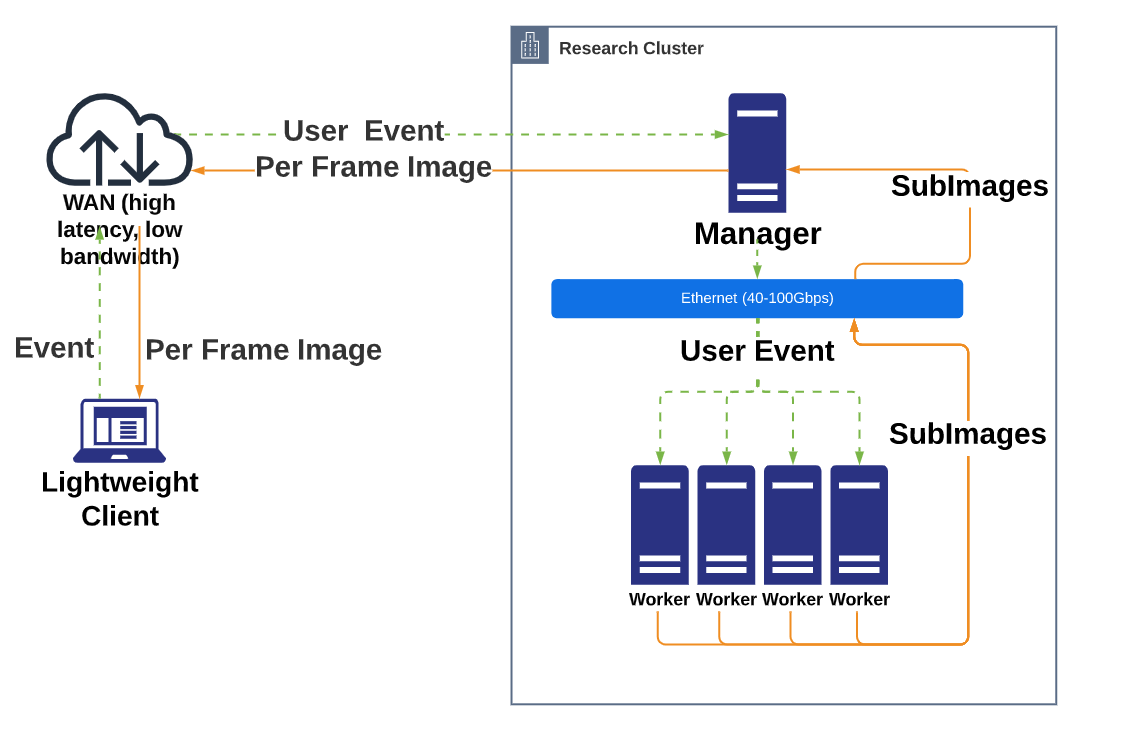} 
\includegraphics[width=0.20\textwidth]{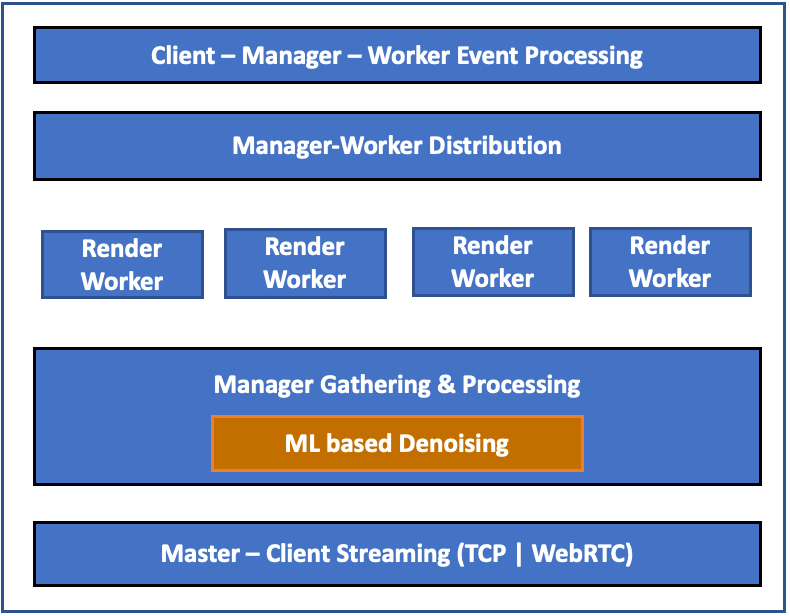}
\caption{Cluster Rendering Architecture}
\label{fig:cluster_architecture}
\vspace{-1.0em}
\end{figure}
\vspace{-0.5em}
\section {Real time Cluster Rendering System}
Since our goal is to enable real time path tracing, 
performance and reliability are the most critical design considerations. 
The ideal cluster rendering system is one in which the performance and quality of 
image generation scales linearly with the number of nodes used, with minimum compute and data transfer overhead.
We present our cluster rendering architecture designed to achieve these goals.
\vspace{-0.5em}
\subsection{Cluster Dataflow Architecture}
We choose a single master multiple worker architecture for its simplicity and minimal data transfer overhead.
In our system, master serves as the central communication hub and the workload manager.
Our design has  $3$ main benefits: 
\begin{itemize}
\item{
It simplified security protection as only the master needs to
support public internet service while workers can remain on a private network.
}  
\item{
Latency overhead introduced by the central master forwarding messages to the workers is negligible $(<1\%)$ compared to the network
latency and network latency variance between client and master.
}
\item {
Latency overhead introduced by the central master merging and processing worker results is significantly less than network,
compute and synchronization overhead required for the client to directly manage workers.
}
\end{itemize}
\vspace{-0.5em}
Figure \ref{fig:cluster_architecture} (left) illustrates the data flow model inside our cluster rendering system, where
each client uses a reliable TCP connection to send UserEvents to the master.
The most basic UserEvent is the CameraEvent which includes the camera pose information for the next frame.
When the master receives the CameraEvent, it immediately forwards it to the workers, each of which 
will perform a portion of the rendering work, then 
send the computed results back to the master.  The master merges all the worker results and performs any post processing
necesary to produce the final rendered image, then sends the final rendered image back to the client for display.

Since intra-cluster network has high bandwidth, low latency and low variance, we use TCP to transmit data between master and workers.
For final image delivery from the master to the client, our system supports both TCP based JPEG image streaming
and UDP based WebRTC video streaming.  
 
\vspace{-0.5em}
\subsection {Render Work Distribution}
To achieve linear scalablity in performance and quality w.r.t. the number of worker nodes
inside a uniform cluster, we want to choose a work distribution strategy with good load balance and minimal overhead in distribution and merging.

Tiling is a common approach to render work distribution where an image of given $width$ and $height$ is divided
into $n$ tiles with fixed tile size. Tiling has small overhead in terms of work
distribution and merging cost; its main drawback is that load balance can be
uneven depending on the scene variation for small tile counts. 

Sample-based distribution leverages the fact that 
path tracing requires many samples per pixel to converge. 
By varying the random seed, each worker can generate a different subset of samples
per pixel.  Each pixel's final radiance is the normalized sum of the radiance 
computed by the workers. 
This approach achieves
perfect load balance by having each worker compute the same number of samples for each pixel.  
However, the network bandwidth required to compute the final radiance
is linear w.r.t. number of worker nodes because every worker needs to send its full resolution radiance buffer
to the master.

\begin{figure}[h]
\includegraphics[width=0.44\textwidth]{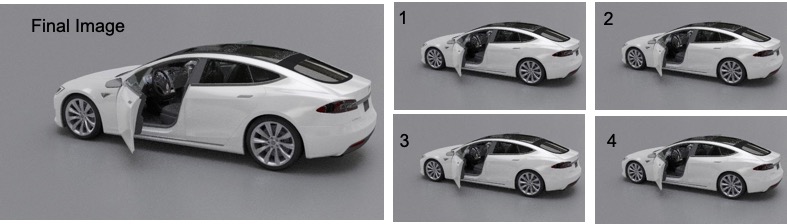}
\includegraphics[width=0.4\textwidth]{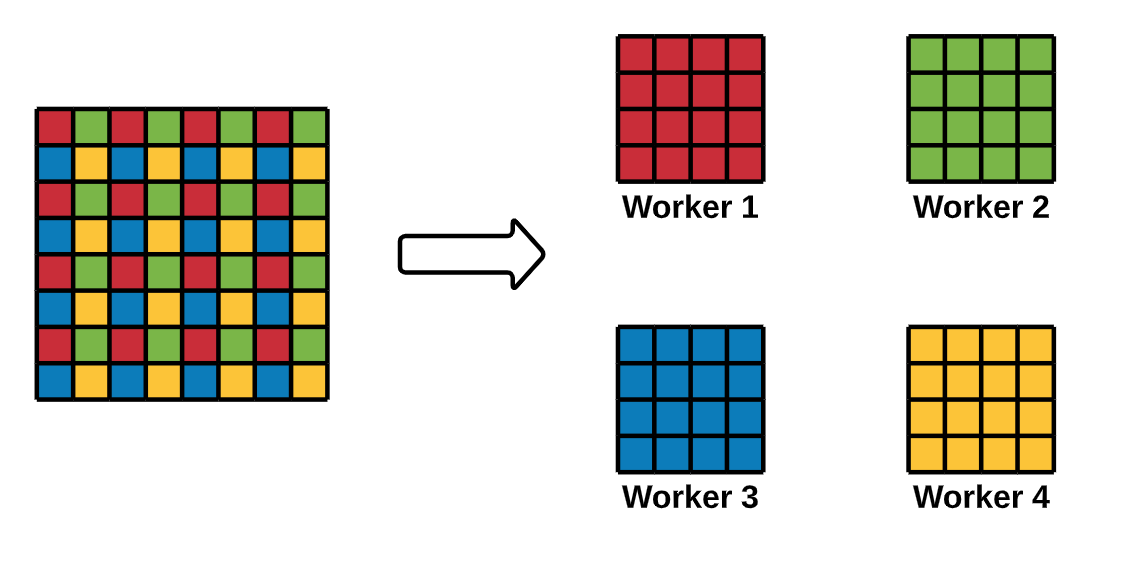}
\vspace{-1em}
\caption{By having each worker render a different sub-sample of the original image, pixel striding delivers near perfect load balance with minimal merging overhead.} 
\label{fig:pixel_stride_icon}
\vspace{-1em}
\end{figure}

Pixel striding divides the rendering work by having each worker render every other pixel.  
For $n$ number of workers, we use a strategy 
where each worker renders a different sub-sample of the original image (Figure~\ref{fig:pixel_stride_icon} top).
Given an  image of size $width$ and $height$,
and the total number of workers being $n = w_n * h_n$,
we ask each worker to generate an image of size 
$\frac{width}{w_n} * \frac{height}{h_n}$, that are combined
together into a final image
with
$\frac{width}{w_n} * \frac{height}{h_n}$ number of pixel blocks of size  $(w_n, h_n)$.

We map each $k$th pixel inside
the $(i, j)$ pixel block of the final image to the $k$th worker's radiance value at pixel $(i, j)$ (Figure~\ref{fig:pixel_stride_icon} bottom).
This is done by 
applying a scale $s = (\frac{1}{w_n}, \frac{1}{h_n})$ to the pixel bounds 
and a translation $t = (s_x\left \lfloor k/w_n\right \rfloor, s_y(k \mod w_n))$  to the pixel center 
by each $k$th worker during per pixel camera-ray sample generation. 

The bandwidth between the worker and the master
is exactly the size of the original radiance buffer and the merge operation for 
all the workers can be done in parallel since each pixel in each worker's output 
corresponds to a different pixel in the final image.
With pixel striding, 
we achieve good load balance 
with minimal compute and bandwidth overhead in work distribution and merging.

\input {pipeline.tex}

%% file: pipeline.tex
\vspace{-0.5em}
\subsection {Pipeline Based Parallel Execution}
 
To achieve maximum 
performance, our cluster rendering system uses pipeline 
based parallel execution models at the highest level, with task level 
parallelism and data parallelism in its compute (path tracing) and 
data (geometry processing) intensive sub-systems where applicable.

Each worker has $2$ main threads:  a main rendering thread that performs
master-forwarded event processing and rendering for frame $n+1$, and
a networking thread that forwards the radiance buffer for frame $n$ to master.

The client also has $2$ main threads:  a UI thread for processing user events and displaying image 
for frame $n-1$,  and a networking thread for receving image (from master) for frame $n$.


The master has $2 + worker\_count$  main threads:
\begin{enumerate}
\item {
A main rendering thread that performs client event processing, work distribution and local rendering for frame $n+1$.
}
\item{
One networking thread per worker for receiving each worker's radiance buffer for frame $n$.
}
\item {
A post processing thread that performs merging of local and worker radiance buffers; followed by denoising, tone mapping and image streaming for frame $n$.
}
\end{enumerate}

Throughout the system, a total of $4 + 3worker\_count$ main threads execute in parallel across 3 pipeline stages. 
To ensure these threads are running at maximum efficiency
without locking or any unnecessary data copy or allocation overhead, fixed sized circular queues are used between each pair of producer and consumer threads.
For example, 
the master's post processing thread is the consumer (reader) of the $local$ RadianceBufferQueue produced (rendered)  by the main rendering thread (writer);
it is also the consumer of the $worker$ RadianceBufferQueue produced by the networking thread (writer); as the master is responsible for merging the radiance
buffers produced by the workers and its own local rendering to produce the complete radiance buffer.

%% file: realtime.tex
\vspace{-0.5em}
\section {Real Time Persistent Cycles}
We choose Cycles (Blender 2.93 version) as the rendering engine for our cluster rendering system
because it has most of the rendering features we need
in geometry (meshes and curves), shading (image based and programmable), physically based
lighting (area lights and environment maps) and BXDF (Disney BSDF, path traced BSSRDF, hair BSDF) models.

Cycles 
has built-in support for Optix-accelerated path tracing and CUDA accelerated
shader evaluation. 
However, being used primarily as an offline production renderer 
where 
each rendered image is treated as completely independent from one another.
It lacks some \emph{basic elements} found in real time
rendering systems, most notably a persistent scene context with the concept of
\emph{frame-time} that can be used to eliminate redoing computation for temporally persistent rendering states.

To enable real time animation rendering,
we need to spend our limited compute resources only on states and geometry 
that \emph{actually} change over time.
First, we added the concept of \emph{frame-time} to the \textit{Scene} object and 
modified Cycles' scene update processing to support \emph{frame-time} notification
to the appropriate nodes and procedurals; then  
we incorporated two experimental features in Blender 3.0, one is support for change registration
for Cycles' node graph, and the other is  
a native Alembic \shortcite{Alembic2015} 
procedural that
enables 
\emph{efficient} playback of prebaked geometry animation stored in this industry standard
format. 

Compared with regular Cycles 2.93,  these changes 
culminated in $9x+$ speed up in per-frame geometry processing for our digital human character
by accelerating 
animation 
playback, geometry data transfer (by only transfering vertex data
that has changed from previous frame), and BVH compute (by using Optix BVH update which is $10x$ more efficient 
than BVH rebuild).
\vspace{-0.5em}
\input{multi_gpu_fig}

\emph{Multiple GPU Performance on a Single Node}.
Cycles uses tiling for render work distribution on a multi-GPU system,  
we use around $60$ tiles to achieve good scaling on our $10$ GPU cluster nodes.
Even though each GPU is responsible for a subset of tiles,
indirect rays in path tracing require 
fast random access to the entire scene and rendering context 
(including the BVH).
High data transfer cost between GPUs means that the most efficient approach is to 
mirror all scene changes, including per frame vertex data
changes and BVH computation.  We replace Cycles' built-in serialized data transfer and serialized BVH compute 
with efficient parallel data transfer and parallel BVH compute. 
Figure \ref{fig:multi_gpu_geom} shows our optimization
changed the multi-GPU scaling curve from linear to (near) constant, and delivered $13x$ (.40s to 0.03s) 
speed up in total geometry processing (data transfer + BVH) and $100x$ speed up (0.32s to 0.003s) 
in BVH computation for the Trudy scene.

%% file: multi_gpu_fig.tex
\begin{figure}[h]
\includegraphics[width=0.225\textwidth]{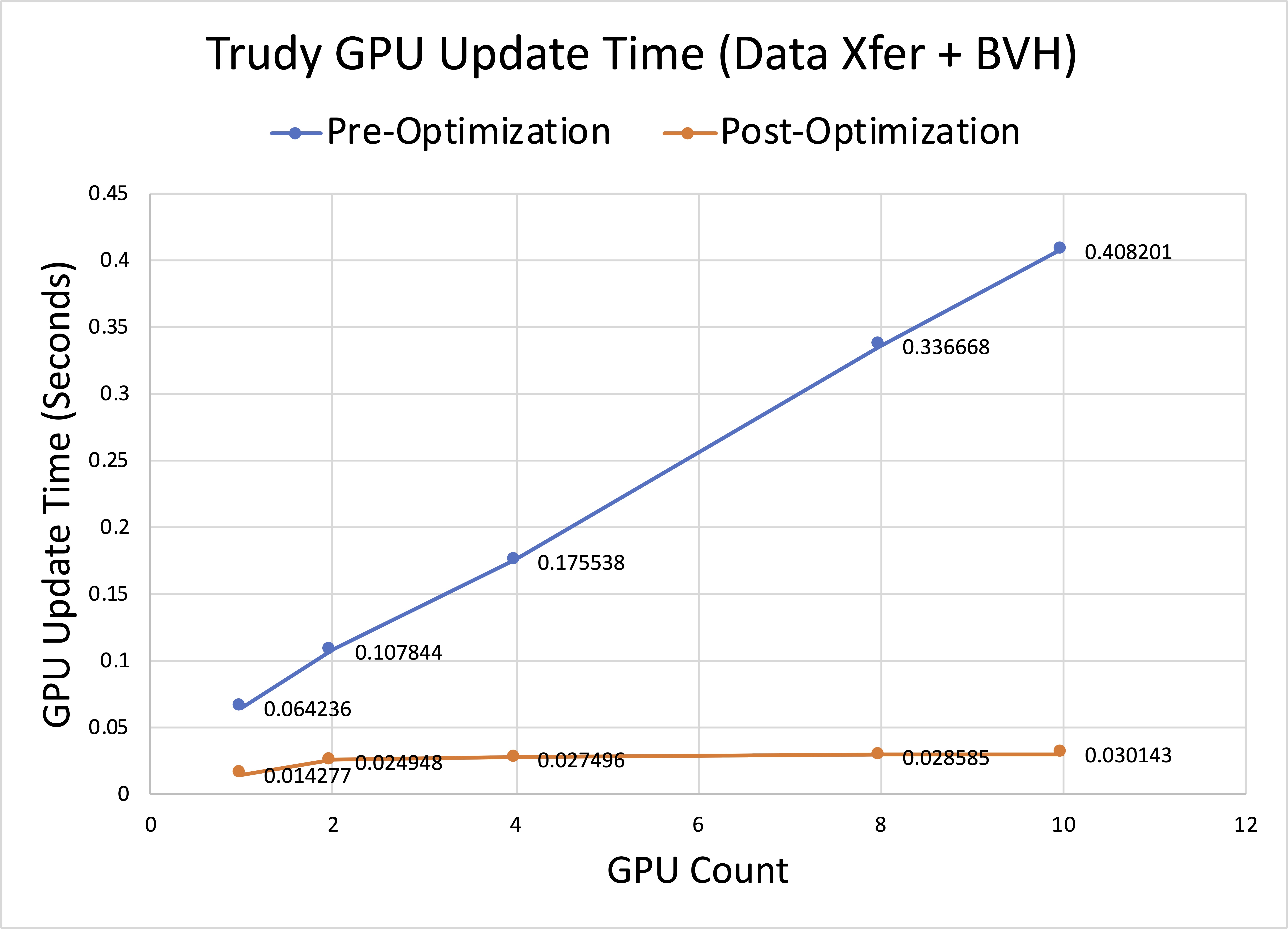}
\includegraphics[width=0.24\textwidth]{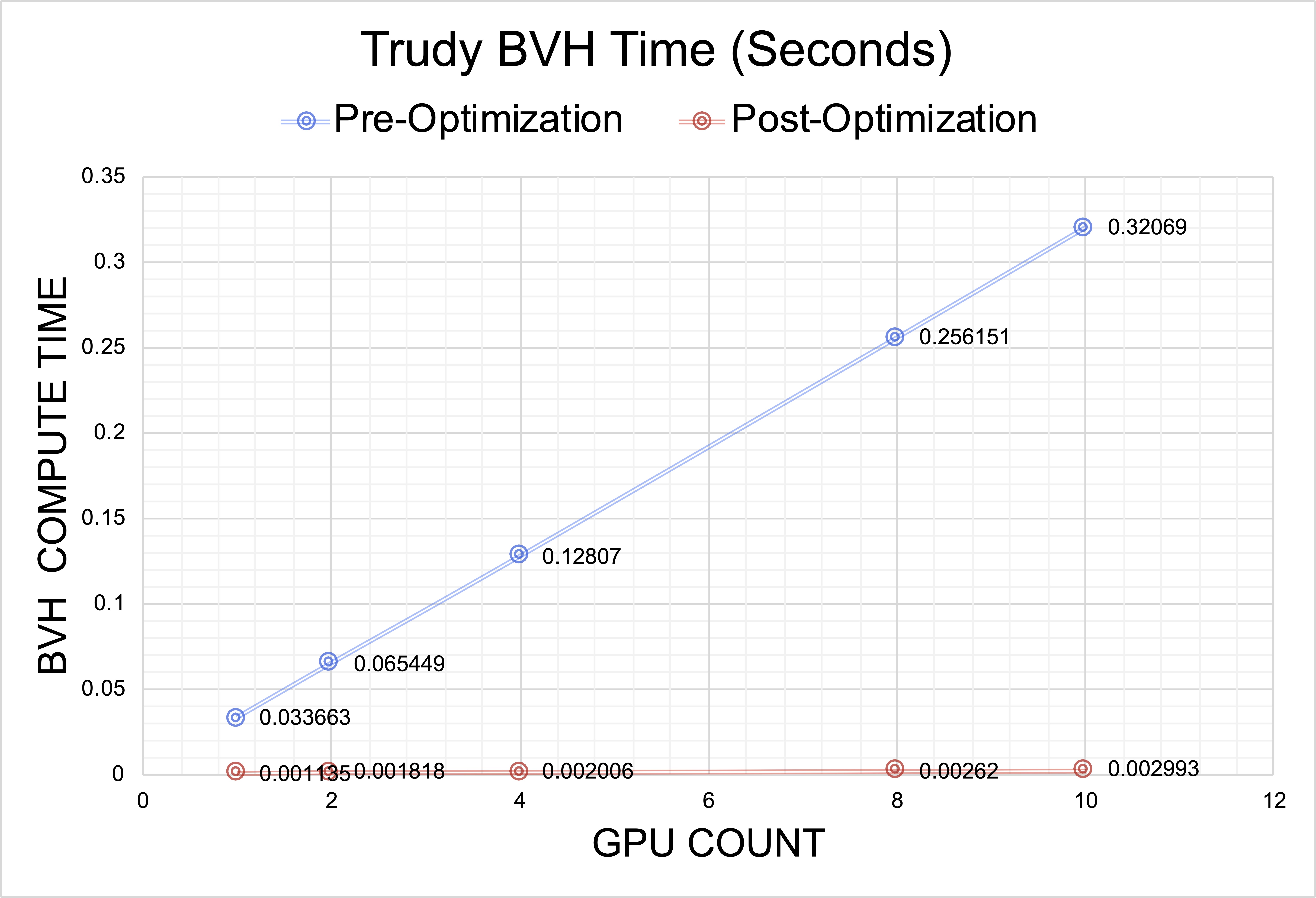}
\caption{Comparison of multi-GPU geometry computation time pre and post  
optimization. 
Left is total GPU geometry update time, right is BVH computation time. 
}
\label{fig:multi_gpu_geom}
\vspace{-0.5em}
\end{figure}

%% file: results.tex
\vspace{-0.5em}
\section {Results}
We present testing results 
on a $32$ node cluster connected with $40$Gb Ethernet \footnote{Node-node latency $\approx 0.1ms$}. Each node
is equipped with $2$ Xeon Gold 6138 (2.00GHz) and $10$ Quadro RTX 8000 ($48$GB PCIe x16).
Every worker and master node in our tests
uses all $10$ GPUs; therefore, $total\_gpu\_used = 10 * nodes\_used$.  

We use two representative scenes
that highlight different rendering effects. Tesla uses glossy reflection and refraction, 
Trudy uses path traced subsurface scatering and physically based
eye shading; both test scenes support inter-object reflections and occlusions with maximum ray depth set to $10$. 

Table~\ref{fig:perf_stats} (left) shows that when total samples per pixel is fixed for Tesla (fixed quality),
increasing the number of working nodes linearly increases frame rate. 
Table~\ref{fig:perf_stats} (right) shows that when frame rate is fixed for Trudy (fixed time), increasing number of working nodes 
linearly improves 
rendering quality (total samples per pixel increases from $50$ to $100$)
\footnote{Rendering quality measured as variance reduction improves as square root of sample count}.

Detailed timing breakdown reported in Table~\ref{fig:perf_stats} shows that our system  
throughput is limited by per frame scene update and render time with $fps \approx \frac{0.9}{update\_time + render\_time}$.
The cluster related overhead of our system (master distribution $+$ merge time) is not only significantly less than rendering time but 
also completely hidden by our pipeline based parallel execution model w.r.t. their impact on rendering throughput (final frame rate).

By making design choices that minimize work distribution overhead and maximize parallel execution efficiency, we have
built the first real-time cluster path tracing renderer that can scale linearly in performance and quality for up to $100$ GPUs.
\input{perfstats}

\vspace{-0.5em}
\section {Conclusion}
We presented the design and implementation of the first production quality
real time cluster path tracing rendering system.  
Our system applies near-optimal work distribution and pipeline based parallel
execution models to deliver almost perfect scaling in path tracing quality and performance 
in a cluster of
RTX nodes connected with high bandwidth interconnect.
However, there remain many optimization and quality improvement opportunites throughout the system and its core
rendering engine that we plan to explore in the future.

%% file: perfstats.tex
\begin{table}[ht]
\begin{tabular}{|c|cc|cc|}
\hline
Scene &\multicolumn{2}{c|}{Tesla} & 
\multicolumn{2}{c|}{Trudy}  \\
\hline
Working Nodes & 2 & 5 & 5 & 10 \\
\hline
Working GPUs  & 20 & 50 & 50 & 100 \\
\hline
Per Node SPP & 15 & 6 & 10 & 10 \\
\hline
Total  SPP 	& 30  & 30 & 50 & 100 \\
\hline
Client FPS  & 15 & 27 & 14 & 14 \\
\hline
Timing Breakdown & \multicolumn{4}{c|} {ms} \\
\hline
Master render & 60.9 & 32.6  & 29.5 & 29.9\\
Worker render  & 59.1 & 26.4 & 34.0 & 39.5\\
Master scene update  & 1.2 & 1.0 & 33.2 & 33.4\\
Worker scene update  & 0.8 & 0.8 & 31.2 & 29.2\\
Master update + render  & 62.1 & 33.6 & 62.7 & 64.3\\
Worker update + render  & 59.9 & 27.2 & 65.2 & 68.7\\
Master tone mapping & 2.0 & 2 & 1.6 & 1.6 \\
Master compression & 8.9 & 8.6 & 11.5 & 9.3\\
Master denoising & 14.7 & 13.9 & 0  & 0\\
\hline
\end{tabular}
\caption {Performance measurements for real-time rendering of Tesla using $2$ and $5$ cluster nodes, 
and of Trudy using $5$ and $10$ cluster nodes. Worker time reported is the average time for all the  workers.}
\vspace{-2em}
\label{fig:perf_stats}
\end{table}

%% file: supplemental_body.tex
\section{Test Scene Description}
The Tesla scene uses $964k$ triangles and $220\ MB$ of unique textures.
The Trudy scene uses $340k$ triangles and $770\ MB$ of unique textures. 
All tests were rendered at the resolution of $1280x720$.

\section{Multi-GPU Rendering Scaling}
\begin{figure}[h]
\includegraphics[width=0.4\textwidth]{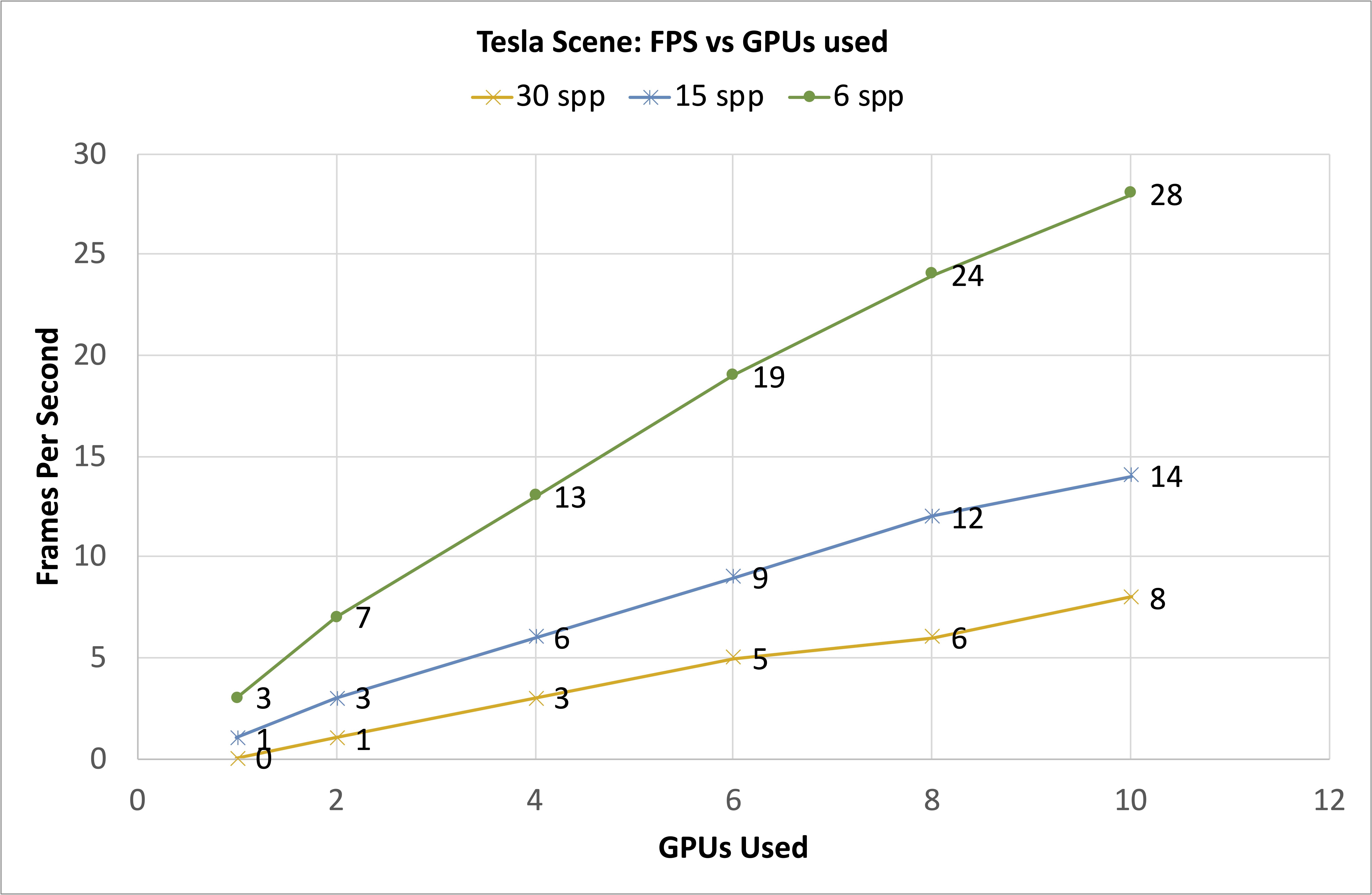}
\caption{Comparison of multi-GPU frame rate change w.r.t. GPUs used.  
}
\label{fig:multi_gpu_render}
\vspace{-0.5em}
\end{figure}

We present measurements of multi-GPU rendering performance on a single cluster node
for the Tesla scene using $3$ different samples per pixel (spp) settings.
Figure~\ref{fig:multi_gpu_render} shows that the measured frame rates increased linearly w.r.t the
number of GPUs used for all $3$ quality settings we tested.

In Table~\ref{tab:perf_stats_tesla}, we present the performance comparison
between rendering Tesla using multiple cluster nodes and
rendering it using a single cluster node.  There is  
no work distribution overhead in the latter case.  We show that the frame rates achieved
when using $1$, $2$ and  $5$ cluster nodes match
closely with the frame rates achieved when rendering Tesla on a single cluster node with $10$ GPUs
at $30$ spp, $15$ spp and $6$ spp (end points on the 3 curves in Figure~\ref{fig:multi_gpu_render}).  
Proving that our cluster rendering system has minimal 
overhead and 
is achieving near optimal performance scaling w.r.t nodes used. 
\input{perfstats_tesla}

%% file: perfstats_tesla.tex
\begin{table}
\begin{tabular}{|c|ccc|}
\hline
Scene &\multicolumn{3}{c|}{Tesla} \\ 
\hline
Working Nodes & 1& 2 & 5   \\
\hline
Worker Count & 0& 1 & 4   \\
\hline
Working GPUs  & 10& 20 & 50  \\
\hline
Per Node SPP & 30 & 15 & 6  \\
\hline
Total  SPP 	& 30 & 30  & 30  \\
\hline
Client FPS  & 8 &  15 & 27  \\
\hline
Timing Breakdown & \multicolumn{3}{c|} {ms} \\
\hline
Master render & 116.3 & 60.9 & 32.6  \\
Worker render  & n/a & 59.1 & 26.4 \\
Master scene update  & 0.6& 1.2 & 1.0 \\
Worker scene update  & n/a & 0.8 & 0.8 \\
Master update + render & 116.9  & 62.1 & 33.6 \\
Worker update + render & n/a  & 59.9 & 27.2 \\
Master tone mapping & 2.1  & 2.0 & 2.0  \\
Master compression & 8.7 & 8.9 & 8.6 \\
Master denoising & 14.5 & 14.7 & 13.9 \\
\hline
\end{tabular}
\caption {Performance measurements for real-time rendering of Tesla using $1$, $2$ and $5$ 
cluster nodes}
\vspace{-2em}
\label{tab:perf_stats_tesla}
\end{table}